\documentclass[twocolumn,letterpaper]{article}
\usepackage[top=0.6in, bottom=1in, left=0.6in, right=0.6in]{geometry}
\usepackage[utf8]{inputenc}
\usepackage{amsmath, amssymb}
\usepackage{graphicx}
\usepackage{subcaption}
\usepackage{authblk}
\usepackage{booktabs}
\usepackage{algorithm}
\usepackage{algorithmic}
\usepackage{url}
\usepackage{threeparttable}

\title{AMAuT: A Flexible and Efficient Multiview Audio Transformer Framework Trained from Scratch \thanks{This preprint version is made available for reference purposes. The final published version may include revisions or formatting changes.}}
\author{Weichuang~Shao}
\author{Iman~Yi Liao}
\author{Tomas~Henrique~Bode~Maul}
\author{Tissa~Chandesa}
\affil{School of Computer and Mathematical Sciences, University of Nottingham Malaysia, Semenyih, Malaysia \\ 
\texttt{andyshao90@gmail.com,\{Iman.Liao, Tomas.Maul, Tissa.Chandesa\}@nottingham.edu.my}}
\date{}

\begin{document}

\maketitle

\begin{abstract}
    Recent foundational models, SSAST, EAT, HuBERT, Qwen-Audio, and Audio Flamingo, achieve top-tier results across standard audio benchmarks but are limited by fixed input rates and durations, hindering their reusability. This paper introduces the \textbf{A}ugmentation-driven \textbf{M}ultiview \textbf{Au}dio \textbf{T}ransformer (AMAuT), a training-from-scratch framework that eliminates the dependency on pre-trained weights while supporting arbitrary sample rates and audio lengths. AMAuT integrates four key components: \emph{(1)} augmentation-driven multiview learning for robustness, \emph{(2)} a conv1 + conv7 + conv1 one-dimensional CNN bottleneck for stable temporal encoding, \emph{(3)} dual CLS + TAL tokens for bidirectional context representation, and \emph{(4)} test-time adaptation/augmentation ($TTA^2$) to improve inference reliability. Experiments on five public benchmarks, AudioMNIST, SpeechCommands V1 \& V2, VocalSound, and CochlScene, show that AMAuT achieves accuracies up to 99.8\% while consuming less than 3\% of the GPU hours required by comparable pre-trained models. Thus, AMAuT presents a highly efficient and flexible alternative to large pre-trained models, making state-of-the-art audio classification accessible in computationally constrained settings.
\end{abstract}

\vspace{1em}
\vspace{1em}

\section{Introduction}
\begin{figure*}[h]
    \centering
    \begin{subfigure}[b]{0.9\textwidth}
        \includegraphics[width=\textwidth]{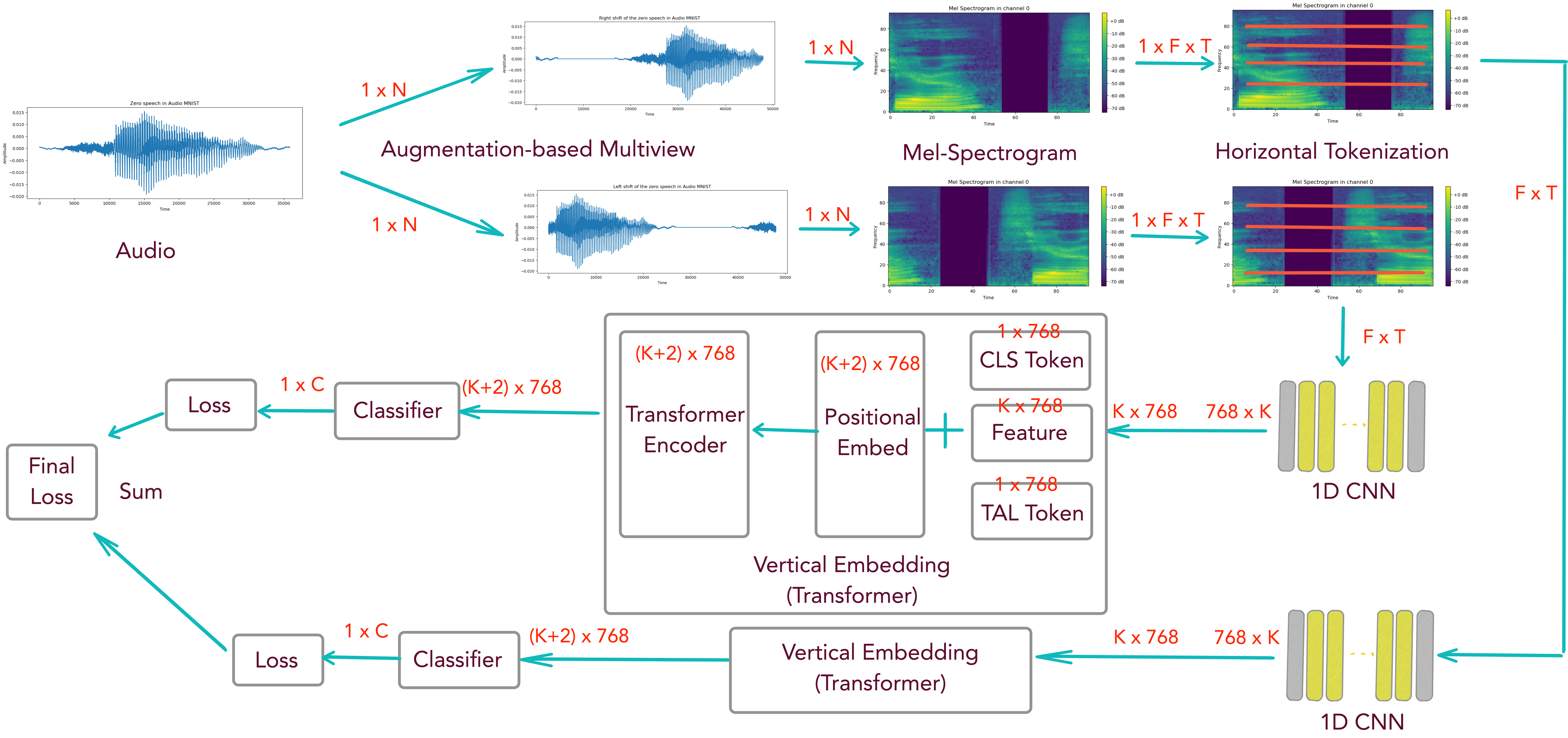}
        \caption{Training and TTDA topology. AMAuT supports more than two views using similar processing steps. Each view processing is independent before combining the loss.}
        \label{Fig:model_arch_f1}
    \end{subfigure}
    \hfill
    \begin{subfigure}[b]{0.85\textwidth}
        \includegraphics[width=\textwidth]{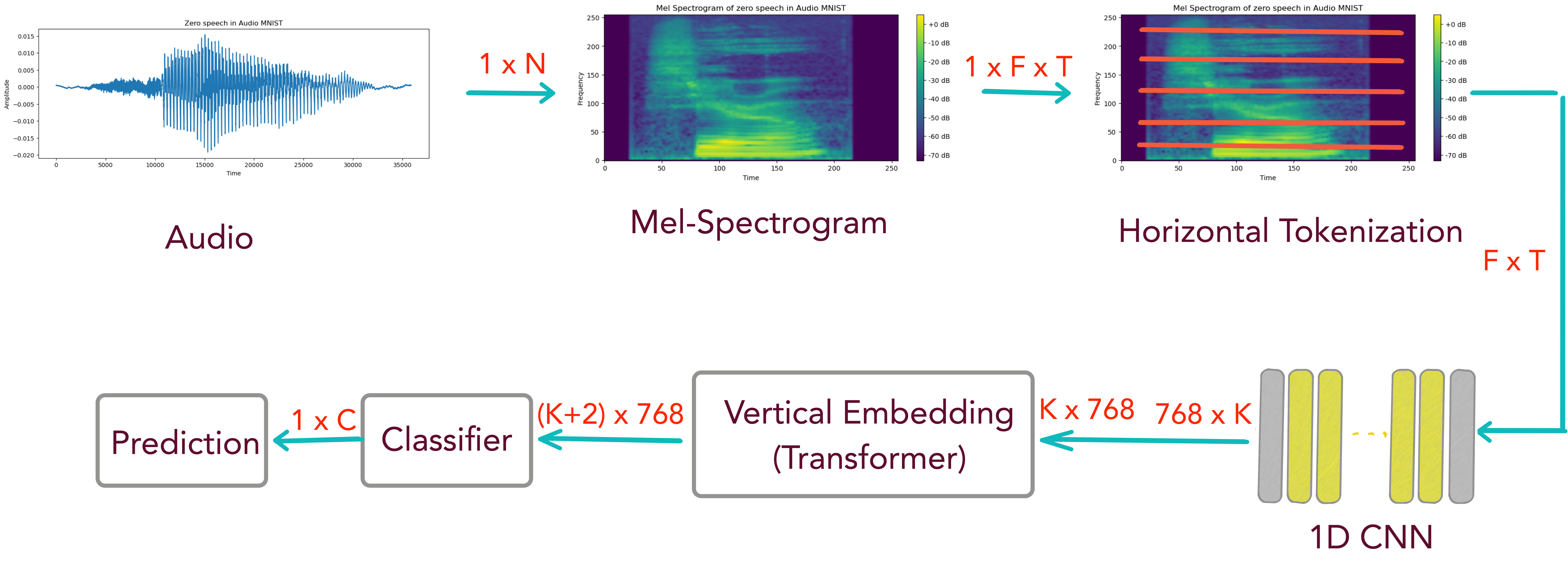}
        \caption{Prediction topology.}
        \label{Fig:model_arch_f2}
    \end{subfigure}
    \caption{Overall architecture of the proposed AMAuT framework. Here, C denotes the number of classes, while N, F, T, and K depend on the dataset.}
    \label{Fig:model_arch}
\end{figure*}

Audio classification is a machine learning task that aims to identify semantic categories from audio signals. Pre-trained models such as SSAST~\cite{gong2022ssast}, EAT~\cite{chen2024eat}, HuBERT~\cite{hsu2021hubert}, Qwen-Audio~\cite{chu2023qwen}, and Audio Flamingo~\cite{kong2024audio}, have achieved state-of-the-art performance on benchmark datasets including SpeechCommands V1 (SC1) \& V2 (SC2)~\cite{warden2018speech}, AudioMNIST (AM)~\cite{audiomnist2023}, VocalSound (VS)~\cite{gong_vocalsound}, and CochlScene (CS)~\cite{jeong2022cochlscene}. However, these models rely heavily on pre-training to reach such performance levels, which imposes several practical limitations. Specifically, reusing pre-trained weights from these models requires a fixed sample rate (16 kHz for SSAST, EAT, HuBERT, and Qwen-audio, and 44.1 kHz for Audio Flamingo) and specific audio length (1 s, 5 s, or 10 s). These constraints restrict model reusability and make it difficult to apply the models to new datasets with different sample rates or audio lengths.

Although resampling techniques can be applied to meet the required rates, such preprocessing often introduces information loss when downsampling or artificial data when upsampling. For instance, \cite{loo2025temporal} reported that the main frequency of bird sounds is below 12 kHz, implying that, based on the information theory~\cite{volkenstein2009entropy} (Nyquist theorem), the sample rate should exceed 24 kHz to preserve information. However, 24 kHz is not a standard rate supported by most existing pre-trained models. Furthermore, as audio recording hardware continues to evolve, producing higher-quality data at higher sample rates, a fixed sample rate methodology cannot take advantage of such an improvement in the recording device.

The same limitation applies to audio length. SSAST and EAT, for example, support only predefined durations of 1, 5, or 10 seconds~\cite{gong2022ssast,chen2024eat}, while Audio Flamingo allows a maximum of 33.25 seconds~\cite{kong2024audio}. When the input audio length or sample rate does not match the model's configuration, pre-training must be repeated or the audio must be forcibly adjusted, both of which are computationally expensive. 

Consequently, this paper addresses a complementary challenge: how to eliminate the fixed input constraints of pre-trained audio transformers. 
To answer this question, this study proposes the \textbf{A}ugmentation-driven \textbf{M}ultiview \textbf{Au}dio \textbf{T}ransformer (AMAuT), a training-from-scratch framework (see Figure~\ref{Fig:model_arch}) that eliminates the dependency on pre-trained weights. This design offers two major advantages: \emph{(1)} AMAuT accepts arbitrary input sample rates and audio lengths, thereby increasing flexibility and adaptability. \emph{(2)} AMAuT significantly reduces computational requirements (GPU hours) by avoiding large-scale pre-training. To maintain performance, AMAuT leverages test-time adaptation/augmentation ($TTA^2$) to enhance inference robustness. 



Major contributions of this paper include:
\begin{enumerate}
    \item \textbf{Augmentation-driven multiview learning}: a training strategy that simultaneously optimizes multiple augmented versions of each audio sample to enhance generalization.
    \item \textbf{Flexible 1D CNN bottleneck}: a conv1 + conv7 + conv1 design that converts variable-length audio inputs into uniform embeddings without resampling.
    \item \textbf{Test-Time adaptation/augmentation ($TTA^2$)}: an inference strategy combining entropy-based domain adaptation and ensemble-style augmentation for improved robustness.
\end{enumerate}

Comprehensive evaluations across five benchmarks reveal that AMAuT achieves accuracy comparable to or surpassing pre-trained counterparts while requiring less than 3\% of their GPU hours. This efficiency highlights the potential of \textbf{pretrain-free}, \textbf{adaptable Transformers} for real-world audio classification where computational budgets are limited.

\section{Related Works}
\subsection{Transformer-based Audio Models}
The transformer architecture, initially developed for language processing, has been effectively adapted for vision and audio tasks. Vision Transformer (ViT)~\cite{dosovitskiy2020image} and Distilled ViT (DeiT)~\cite{touvron2021training} operate by dividing images into fixed-size patches (e.g., $16\times 16$ or $32\times 32$), a concept later transferred to the audio domain, such as Audio Spectrogram Transformer (AST)~\cite{gong2021ast}.
AST represents audio as Mel-spectrograms~\cite{ustubioglu2023mel,hwang2020mel}, turning temporal signals into two-dimensional time–frequency grids before applying patch-based Transformer processing with fixed-size patches ($16 \times 16$).
While AST reuses ImageNet-trained weights~\cite{deng2009imagenet,gong2021ast}, SSAST~\cite{gong2022ssast} processes pre-training on AudioSet~\cite{gemmeke2017audio} and LibriSpeech~\cite{panayotov2015librispeech} rather than reusing ImageNet-trained weights for extending AST. 

Models such as Wav2Vec 2.0~\cite{baevski2020wav2vec} and HuBERT~\cite{devlin2019bert} process raw wavforms using 1D CNN-Transformer architectures. Wav2Vec 2.0 is composed of seven convolutional layers equipped with LayerNorm and GELU activations, and its large-scale pre-training on LibriSpeech~\cite{panayotov2015librispeech} and Libri-Light~\cite{kahn2020libri} reportedly required about 7.5 days on 128 GPUs.
HuBERT follows a similar CNN-Transformer architecture to Wav2Vec 2.0 but extends it with a BERT-style~\cite{devlin2019bert} masked prediction mechanism that integrates a waveform encoder.
Like Wav2Vec 2.0, AMAuT adopts a 1D CNN-Transformer structure but differs fundamentally: AMAuT trains from scratch without pre-training, uses 1D CNNs on Mel-spectrogram horizontal tokens, and replaces LayerNorm + GELU with BatchNorm + ReLU. These changes improve both efficiency and accuracy (99.82\% vs. 99.4\% on AudioMNIST) between Wav2Vec 2.0 and AMAuT.

The spectrogram transformers~\cite{zhang2022spectrogram} contributed a Tempo-Frequency sequential (TFS) attention for separating mel-spectrogram outputs within frequency direction instead of $16 \times 16$ regions in AST~\cite{gong2021ast} that are similar to the horizontal tokenization of AMAuT (see subsection~\ref{ssec:horizontal_tokenization}). However, \cite{zhang2022spectrogram} inherits the methodology of AST and ViT, and then \cite{zhang2022spectrogram} employs a different model architecture, as it incorporates 2D CNN layers.

Overall, both AST and SSAST constrain input to fixed sample rates (typically 16 kHz) and durations (1 s, 5 s, or 10 s) when reusing their pre-trained weights; Wav2Vec 2.0 and HuBERT are computationally expensive and also fixed at 16 kHz. AMAuT removes these constraints, allowing flexible sample rates and audio lengths while matching or exceeding the accuracy of pre-trained models at a cost that is at most 3\% of their GPU hours.

\subsection{Test-Time Domain Adaptation}
Test-Time domain adaptation (TTDA) only uses the test set for performance improvement by adapting (partial) model parameters of a trained model from a source domain to a target domain~\cite{liang2024comprehensive}. In AMAuT, this is achieved through a composite loss function $\mathcal{L}_{TTDA}$ that combines three complementary objectives: Nuclear-Norm Maximization, Entropy Minimization, and Generalized Entropy (see Eq.\ref{eq:nm_loss}, Eq.\ref{eq:entropy_loss}, and Eq.\ref{eq:g-entropy_loss} in Appendix~\ref{app:algorithm}). These components have been applied separately in previous studies, such as CoNMix~\cite{kumar2023conmix} for test-time domain adaptation, TENT~\cite{wang2021tent} for entropy-based test-time batch adaptation, and SGEM~\cite{kim2023sgem} for Automatic Speech Recognition, but AMAuT unifies them for the first time. This integration allows efficient test-time domain adaptation and improves robustness under moderate distribution shifts without additional supervision.

\section{Methodology}
\subsection{Overview}
The proposed Augmentation-driven Multiview Audio Transformer (AMAuT) architecture (see Figure~\ref{Fig:model_arch}) follows a hybrid CNN-Transformer pipeline designed for flexibility and efficiency. Unlike prior models constrained by pre-trained feature extractors, AMAuT is trained entirely from scratch and can ingest audio at \textbf{arbitrary sampling rates and lengths}.

During training, each audio sample undergoes \textbf{multiview augmentation}, producing several perturbed variants (time-shifted, noise-added, or background-mixed). These parallel views are passed through identical processing streams, and their corresponding losses are jointly optimized to encourage invariance to perturbations.

Each augmented view is first converted to a \textbf{Mel-spectrogram}, then segmented along the temporal dimension through \textbf{horizontal tokenization}. A \textbf{1D CNN} transforms these tokens into a fixed-size representation (768 × K), which captures fine-grained temporal dependencies with reduced computational cost compared to 2D CNNs.

The resulting embeddings enter the \textbf{vertical embedding transformer}, including CLS + TAL tokens and learnable positional embeddings. Multi-head self-attention layers refine the token relationships before classification.

During adaptation, AMAuT applies $TTA^2$, a lightweight refinement stage combining (a) test-time domain adaptation (TTDA) that updates internal statistics using unlabeled test data, and (b) test-time augmentation (TTAu) by averaging predictions across multiple augmented inputs. Together, these mechanisms improve robustness to unseen conditions without additional training labels.

\subsection{Multiview Augmentation} 
During both training and test-time domain adaptation (TTDA) (see Figure~\ref{Fig:model_arch_f1}), the AMAuT framework employs \textbf{augmentation-driven multiview learning} to improve model robustness.
 
Multiple augmented views are generated from each input sample, and each view contributes to the overall loss calculation (see Algorithm~\ref{alg:aug_multiview} in Appendix~\ref{app:algorithm}). Rather than computing gradients for each loss individually, AMAuT aggregates all losses before performing gradient updates:
\begin{equation}\label{eq:multiview_loss}
    \sum^A_i \mathcal{L}(f(x_i;\theta))
\end{equation}
where $\mathcal{L}$ denotes the loss function, $f(\cdot, \theta)$ represents the AMAuT model, $A$ denotes the number of augmentation-driven views, and $\{x_i|1\le i \le A\}$ are the augmented views of the same audio sample $x$.

During validation, AMAuT uses the original audio input $x$ without multiview augmentation to ensure a fair comparison with other state-of-the-art models (see Figure~\ref{Fig:model_arch_f2}).

In the experiments, four augmented views were used during training and two during TTDA (see Subsection~\ref{ssec:tta}). All augmentations introduce random variations, such as time shifts (Figure~\ref{Fig:model_arch_f1} demonstrates the left and right time-shift augmentations) and Gaussian or background noise, ensuring that each epoch encounters unique augmented samples. 

Specifically:
\begin{itemize}
    \item \textbf{Training} (four augmentations): \emph{(1)} random left or right time shift (0-17\%), \emph{(2)} Gaussian noise (noise-to-audio ratio = 0.015, and regenerate noise each time), and \emph{(3-4)} random clips from two background noises (“dude miaowing” and “pick noise” from SpeechCommands~\cite{warden2018speech}) with an SNR of 50 dB.
    \item \textbf{TTDA} (two augmentations): \emph{(1)} random right time shift (0-17\%), and \emph{(2)} random left time shift (0-17\%).
\end{itemize}

\subsection{Spectrogram and Horizontal Tokenization}\label{ssec:horizontal_tokenization}
Before tokenization, each audio sample is transformed into a \textbf{2D Mel-spectrogram}, where the height and width correspond to frequency and time dimensions, respectively. AMAuT processes only \textbf{mono audio}. For multichannel audio, it converts the signals to mono by averaging across channels as follows:
\begin{equation}\label{eq:ster2mono}
    \begin{split}
        \text{sterToMono}(x) = \{\frac{1}{\mathcal{C}}\sum_{i=1}^\mathcal{C} x_{i,j}|1\le j \le N\}\\
    \end{split}
\end{equation}
where $x = \{x_{i,j}|1\le i \le \mathcal{C}, 1\le j \le N\}$ is a multi-channel audio signal, and $\text{sterToMono}(x)$ is the resulting mono signal. Here, $\mathcal{C}$ represents the number of channels and $N$ is the length of each channel. 

After conversion to mono, the spectrogram has a shape of $1 \times F \times T$, where $F$ and $T$ denote the frequency and time dimensions, respectively. The channel dimension (1) is removed, reshaping the spectrogram to $F \times T$ for subsequent tokenization (see Figure~\ref{Fig:model_arch}).

Horizontal tokenization divides the spectrogram along the time dimension, producing a sequence of \textbf{temporal tokens}. These tokens serve as input to the subsequent 1D CNN module, which efficiently captures temporal dependencies.

\subsection{1D CNNs}
The architecture uses \textbf{1D CNNs} because it interprets the spectrogram's frequency bins as feature channels, leveraging the independent nature of different frequencies in an audio signal. Further, 1D CNNs consume less GPU memory and computation cost, as a 1D convolution (conv7) has a smaller kernel size than a 2D convolution (conv3×3). The latter requires two-dimensional kernels that significantly increase GPU memory usage. The 1D CNN module processes the horizontal tokens and reshapes them into a consistent feature size of $768 \times K$ (see Figure~\ref{Fig:model_arch}).

\begin{table}[h]
  \centering
  \caption{List the AMAuT configurations for different datasets, including the number of convolutional layers, GPU memory usage, and total parameters.}
  \label{tab:AuT_traits}
  \begin{tabular}{ccccl}
    \toprule
    Set & CNN Depth & GPU Memory & Parameters \\
    \midrule
    AM & 49 & 2.63 GB & 99.1 M \\
    SC1 & 61 & 4.79 GB & 103.1 M \\
    SC2 & 61 & 4.20 GB & 103.1 M \\
    VS & 64 & 9.47 GB & 97.0 M \\
    CS & 79 & 12.89 GB & 101.0 M \\
    \bottomrule
  \end{tabular}
\end{table}

Longer audio inputs require deeper CNN stacks to reduce temporal resolution to 768 tokens, resulting in dataset-specific CNN configurations (see Table~\ref{tab:AuT_traits}).
\begin{figure}[h]
    \centering
    \includegraphics[width=0.5\linewidth]{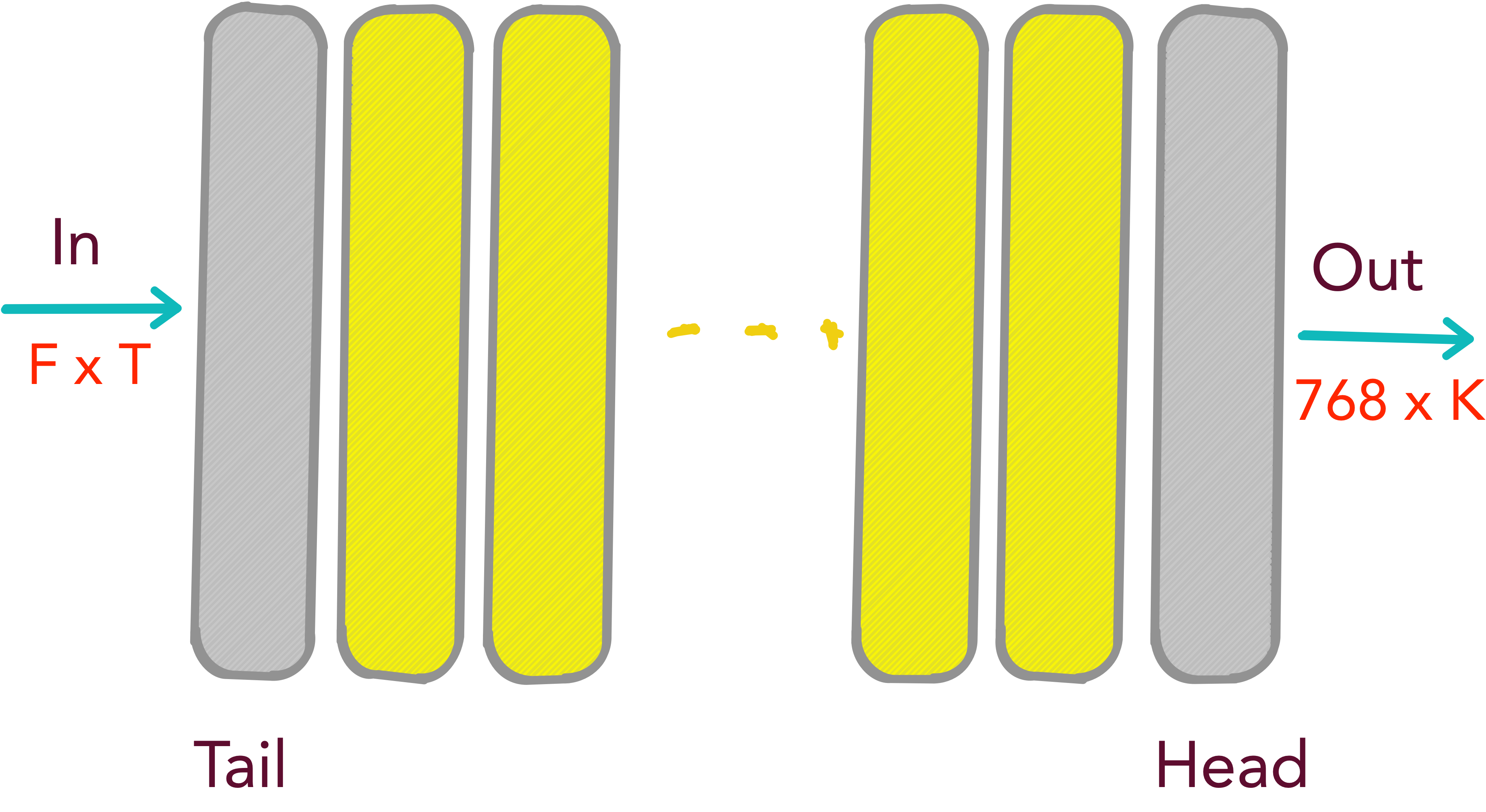}
    \caption{Structure of the 1D CNN module used in AMAuT}
    \label{fig:1D_cnn}
\end{figure}
As illustrated in Figure~\ref{fig:1D_cnn}, the 1D CNN module consists of:
\begin{itemize}
    \item \textbf{Tail block}: a conv14 + max-pooling layer for dimensionality reduction;
    \item \textbf{Bottleneck blocks (yellow)}: multiple conv1 + conv7 + conv1 sequence blocks for feature extraction;
    \item \textbf{Head block}: a final conv1 layer that fixes the output shape to be $768 \times K$.
\end{itemize}
This architecture efficiently extracts temporal representations while maintaining manageable computational complexity, making it suitable for diverse audio lengths and sample rates.

\subsection{Vertical Embedding}
\begin{figure}[h]
    \centering
    \includegraphics[width=0.8\linewidth]{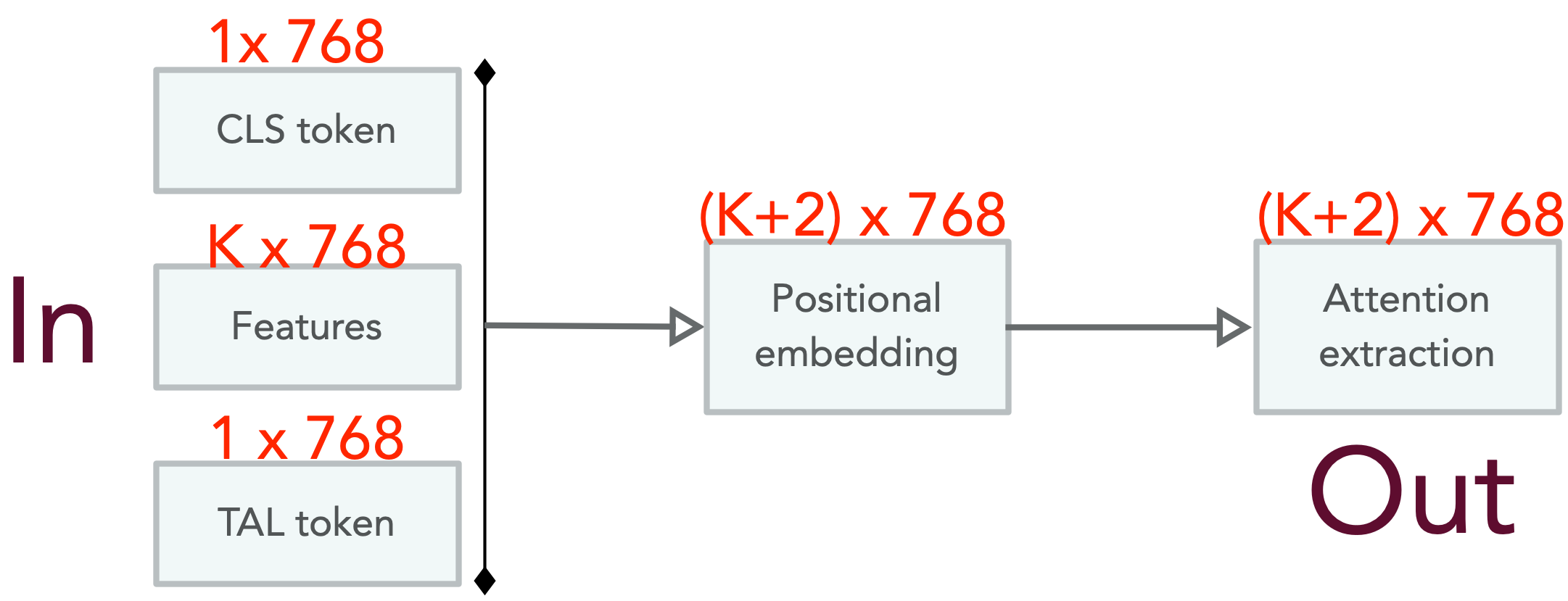}
    \caption{Architecture of the Vertical Embedding module in AMAuT}
    \label{Fig:model_arch_f3}
\end{figure}
As shown in Figure~\ref{Fig:model_arch_f3}, the vertical embedding module receives 1D CNN features with a shape of $768 \times K$. Since vertical embedding converts the tensor to $K \times 768$ by \textbf{transpose}, the process is referred to as \textit{vertical embedding}.

\subsubsection*{CLS+TAL Tokens and Learnable Positional Embedding} 
AMAuT introduces two learnable tokens: the \textbf{class token (CLS)} and the \textbf{tail token (TAL)} (see Eq.~\ref{eq:CLS_token} and Figure~\ref{Fig:model_arch_f3}). Similar to DeiT~\cite{touvron2021training}, AMAuT positions CLS and TAL at opposite ends of the token sequence. 
 
Before the attention extraction stage, AMAuT applies a \textbf{learnable positional embedding} and a \textbf{random dropout layer} (15\% drop rate) to the concatenated token sequence:
\begin{align}
    h & = \text{concate}(\text{CLS}, x, \text{TAL}) \label{eq:CLS_token} \\
    \hat{x} & = \text{mask}(h + \mathcal{P})
\end{align}
where $x$ denotes the tokenized features, $\hat{x}$ the embedding input, $\mathcal{P} = \{\mathcal{P}_0, \cdots, \mathcal{P}_{768+1}\}$ the positional embedding table, and $\text{mask}(\cdot)$ the dropout layer.

\subsubsection*{Attention Extraction} 
This component follows the Transformer~\cite{vaswani2017attention} architecture and comprises multiple transformer blocks. 
Two important configurations are adopted:
\begin{enumerate}
    \item The dropout rate for both multi-head attention and MLP is set to 0\%, consistent with models such as SSAST~\cite{gong2022ssast}, AST~\cite{gong2021ast}, and CoNMix~\cite{kumar2023conmix}.
    \item AMAuT uses a \textbf{full attention matrix} rather than a lower-triangular one, as it performs classification rather than generative prediction. This choice enhances performance by preserving all temporal dependencies.
\end{enumerate}

\subsection{Classifier}
\begin{figure}[h]
    \centering
    \begin{subfigure}[b]{0.45\textwidth}
        \includegraphics[width=1.0\textwidth]{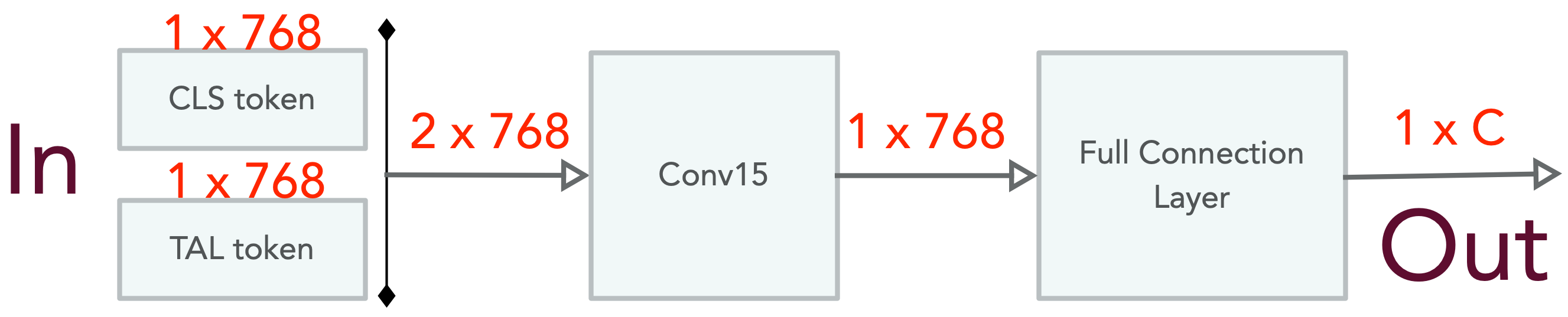}
        \caption{ }
        \label{fig:classifier_f1}
    \end{subfigure}
    \begin{subfigure}[b]{0.45\textwidth}
        \includegraphics[width=1.0\textwidth]{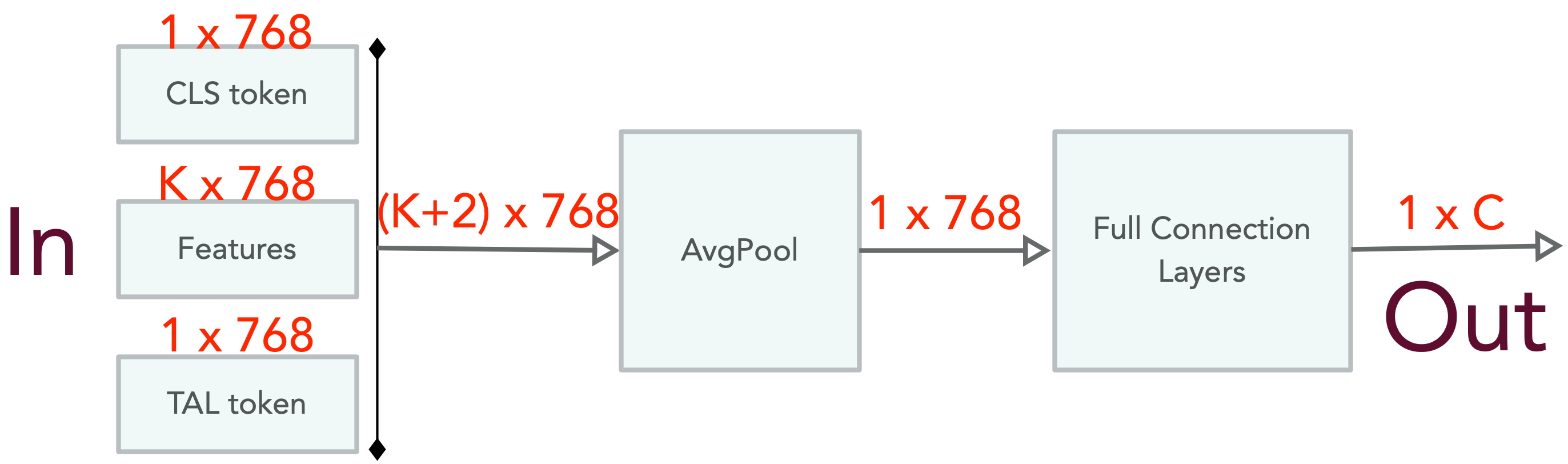}
        \caption{ }
        \label{fig:classifier_f2}
    \end{subfigure}
    \caption{Classifier architectures used in AMAuT for different audio durations. (a) For audio $>$1s, only CLS + TAL tokens are used, followed by a 1D CNN and a fully connected (FC) layer. (b) For 1s inputs, using all features with AvgPool and FC layers.}
    \label{fig:classifier}
\end{figure}
As illustrated in Figure~\ref {fig:classifier}, AMAuT employs two classifier architectures based on the input audio duration.
\begin{itemize}
    \item \textbf{For inputs longer than 1 second (Figure~\ref{fig:classifier_f1})}: \\
    Only the CLS and TAL tokens are used. The classifier comprises one 1D CNN layer followed by a fully connected (FC) layer that maps features to class probabilities.
    \item \textbf{For 1-second inputs (Figure~\ref{fig:classifier_f2})}: \\
    All transformer-extracted features are used. The classifier applies average pooling, followed by three FC layers with batch normalization, reducing the dimensionality to the number of output classes.
\end{itemize}
This dual-strategy design ensures efficient processing for short inputs while maintaining representational richness for longer sequences.


\subsection{Test-Time Adaptation / Augmentation ($TTA^2$)}\label{ssec:tta}
At inference, AMAuT employs $TTA^2$, combining two complementary strategies:
 
\subsubsection*{Test-Time Doman Adaptation (TTDA)}
This subsection investigates an uncommon application of Test-time Domain Adaptation (TTDA): improving accuracy when training and test distributions are similar. Typically, TTDA is employed to address distribution shifts, but we explore its utility even when such a shift is minimal~\cite{liang2024comprehensive}.

TTDA enhances performance by adapting a trained model to the test data distribution without requiring labels~\cite{liang2024comprehensive}. During TTDA, all test features are visible, but labels remain hidden. The model updates its internal parameters to better align with the target domain before final prediction.

TTDA follows the same computational structure as training but employs a different loss $\mathcal{L}_{TTDA}$ rather than the training loss $\mathcal{L}_{Tr}$ (see Figure~\ref{Fig:model_arch_f1}). AMAuT combines three entropy-based losses into a single objective (see Eq.\ref{eq:nm_loss}, Eq.\ref{eq:entropy_loss}, and Eq.\ref{eq:g-entropy_loss} in Appendix~\ref{app:algorithm}):
\begin{equation}\label{eq:ttl_tta_loss}
    \mathcal{L}_{TTDA} = \alpha\mathcal{L}_{NM} + \beta\mathcal{L}_{EN} + \gamma\mathcal{L}_{GEN}
\end{equation}
where $\alpha$, $\beta$, and $\gamma$ are scaling coefficients (see Table\ref{tab:ttda_param} in Appedix~\ref{app:param_set}).

By integrating these complementary objectives, AMAuT improves inference robustness in scenarios where training and testing data share similar but not identical distributions.

\subsubsection*{Test-Time Augmentation (TTAu)}
To further enhance inference robustness, AMAuT adopts test-time augmentation (TTAu)~\cite{kimura2021understanding}. Unlike TTDA, TTAu does not update model parameters; instead, it refines predictions by averaging outputs from multiple augmented versions of the same input. Three TTAu strategies are evaluated: Augmentation-driven refinement (Aug), Multi-training-based refinement (Mlt), and Hybrid refinement (Hyb) (see Eq.\ref{eq:aug_rf}, Eq.\ref{eq:mlt_rf}, and Eq.\ref{eq:hyb_rf} in Appendix~\ref{app:algorithm}, separately).

\subsection{Learning Rate and Optimization}
Both the training loss $\mathcal{L}_{Tr}$ (see Eq.~\ref{eq:train_loss} in Appendix~\ref{app:algorithm}) and TTDA loss $\mathcal{L}_{TTDA}$ (see Eq.~\ref{eq:ttl_tta_loss}) employ a \textbf{dynamic learning rate scheduler} to accelerate convergence. The scheduler begins with a relatively large learning rate and gradually reduces it as training progresses, ensuring stable optimization. The update rule is defined as
\begin{align}\label{eq:lr_scheduler}
    lr_e = \begin{cases}
        lr_{e-1} (1 + \lambda \cdot e / \eta)^{0.75}, & \text{if } e < \eta \\
        lr_{e-1}, & \text{otherwise}
    \end{cases}
\end{align}
where $e$ denotes the epoch number, $10 \le \lambda \le 40$ controls the rate of change, $200 \ge \eta \ge 40$ defines the cardinality. See more details of hyperparameter settings from Table~\ref{tab:LRS_param} in Appendix~\ref{app:param_set}.

During both training and TTDA, AMAuT uses the \textbf{stochastic gradient descent (SGD)} optimizer with weight decay $10^{-3}$, momentum 0.9, and Nesterov acceleration enabled. 

\section{Experiments and Results}
AMAuT integrates several architectural and training strategies, including augmentation-driven multiview learning, conv1 + conv7 + conv1 CNN bottlenecks, flexible handling of arbitrary sample rates and audio lengths, learnable positional embeddings, and $TTA^2$. Collectively, these components enable AMAuT to achieve strong performance when trained entirely from scratch, without relying on large-scale pre-training.
\subsection{Datasets}
\begin{table}[h]
    \centering
    \caption{Dataset Properties}
    \label{tab:dataset}
    \begin{tabular}{cccl}
        \toprule
        Set & Size & Length & Rate \\
        \midrule
        AM & 30000 & 1 s & 48 kHz \\
        SC1 & 64721 & 1 s & 16 kHz \\
        SC2 & 105829 & 1 s & 16 kHz \\
        VS & 20977 & 12 s & 16 kHz \\
        CS & 76115 & 10 s & 44.1 kHz \\
        \bottomrule
    \end{tabular}
\end{table}
To evaluate the proposed AMAuT framework, this study employed five publicly available benchmark datasets: AudioMNIST (AM)~\cite{audiomnist2023}, SpeechCommands V1 (SC1) \& V2 (SC2)~\cite{warden2018speech}, VocalSound (VS)~\cite{gong_vocalsound}, and CochlScene (CS)~\cite{jeong2022cochlscene}.

As shown in Table~\ref{tab:dataset}, these datasets cover a diverse range of sampling rates and audio durations, enabling a comprehensive analysis of AMAuT’s flexibility and robustness. Specifically, AM operates at \textbf{48 kHz}, CS at \textbf{44.1 kHz}, and the remaining datasets at \textbf{16 kHz}. Additionally, audio lengths also vary substantially, including \textbf{12 s} for VS, \textbf{10 s} for CS, and \textbf{1s} for the others, allowing AMAuT to be evaluated on both short and long clips. Thus, this experimental configuration enabled consistent comparisons across all datasets.

For model training, AMAuT used the \textbf{training and validation splits} for AM, and the \textbf{training sets} for SC1, SC2, VS, and CS.

\subsection{Comparative Performance}
Table~\ref{tab:accu_cmp} compares AMAuT with several state-of-the-art pre-trained models across the five datasets.
\begin{table}[ht]
    \centering
    \caption{Comparison of prediction accuracy (\%) between the proposed AMAuT and state-of-the-art pre-trained models on five benchmark datasets.}
    \label{tab:accu_cmp}
    \begin{threeparttable}
    \begin{tabular*}{.5\textwidth}{@{\extracolsep\fill}ccl}
        \toprule
        Algorithm & Set & Accuracy(\%) \\
        \midrule
        AMAuT & SC1 & \textbf{96.31} \\
        SSAST~\cite{gong2022ssast} & SC1 & 96.20 \\
        \midrule
        AMAuT & SC2 & 96.21 \\
        EAT~\cite{chen2024eat} & SC2 & \textbf{98.30} \\
        AST~\cite{gong2021ast} & SC2 & 98.11 \\
        PDC~\cite{chrysos2022augmenting} & SC2 & 97.8 \\
        \midrule
        AMAuT & AM & 99.82 \\
        HuBERT~\cite{hsu2021hubert,la2024benchmarking} & AM & \textbf{99.95} \\
        XLS-R~\cite{babu2021xls,la2024benchmarking} & AM & 99.86 \\
        Wav2Vec2~\cite{baevski2020wav2vec, la2024benchmarking} & AM & 99.42 \\
        \midrule
        AMAuT & VS & 92.26 \\
        Qwen-Audio~\cite{chu2023qwen} & VS & \textbf{92.89} \\
        EfficientNet~\cite{gong_vocalsound,tan2019efficientnet} & VS & 90.5 \\
        \midrule
        AMAuT & CS & 72.58 \\
        Audio Flamingo~\cite{kong2024audio} & CS & \textbf{83.0} \\
        Qwen-Audio~\cite{chu2023qwen} & CS & 79.5 \\
        \bottomrule
    \end{tabular*}
    \begin{tablenotes}
        \item[1] Performance values for baseline models are obtained from their respective original publications.
    \end{tablenotes}
    \end{threeparttable}
\end{table} 
AMAuT achieved \textbf{96.31\%} on SC1 and \textbf{99.82\%} on AM, closely matching or exceeding prior results, while maintaining minimal computational overhead.

On VS, AMAuT’s accuracy (\textbf{92.26\%}) was only 0.63\% below Qwen-Audio [9], and it remained competitive even on the challenging CS dataset (\textbf{72.58\%}), where pre-trained models still hold an advantage.

The results suggest that, with an adaptable design, AMAuT trained entirely from scratch can approach or match pre-trained performance in many scenarios.

\subsection{Ablation Studies}
\subsubsection*{Effectiveness of Multiview Augmentation}
The experiments support the claim that \textbf{augmentation-driven multiview learning} provides superior performance to conventional augmentation methods. Unlike conventional sequential or single-view augmentation, the proposed multiview approach applies multiple random augmentations to each input sample during both training and test-time domain adaptation (TTDA). Rather than generating and storing augmented data offline, AMAuT performs augmentations in real time, enabling transformations to vary across epochs. Consequently, the model is continually exposed to diverse variants of the same source sample, improving generalization and robustness.
\begin{table}[h]
    \centering
    \caption{Comparison of test accuracies between sequential augmentation and augmentation-driven multiview learning on CS and SC1 before applying $TTA^2$.}
    \label{tab:seq_mlt_cmp}
    \begin{threeparttable}
    \begin{tabular*}{.5\textwidth}{@{\extracolsep\fill}cccl}
        \toprule
        Set & Aug-method & 1-Aug(\%) & 4-Aug(\%) \\
        \midrule
        CS & sequence & 65.37 & 41.46$\downarrow$ \\
        CS & multiview & 65.37 & \textbf{68.87}$\uparrow$ \\
        \midrule
        SC1 & sequence & 95.26 & 93.01$\downarrow$ \\
        SC1 & multiview & 95.26 & \textbf{95.93}$\uparrow$ \\
        \bottomrule
    \end{tabular*}
    \begin{tablenotes}
        \item[1] “1-Aug” and “4-Aug” indicate the use of one or four augmented views, respectively.
        \item[2] The arrows ($\uparrow$, $\downarrow$) denote performance improvement or degradation relative to the baseline.
    \end{tablenotes}
    \end{threeparttable}
\end{table} 

Compared with the traditional dataset-extension method (offline), where augmented samples are pre-generated and stored for later training, the augmentation-driven multiview learning approach offers two significant advantages:
\begin{enumerate}
    \item \textbf{Dynamic sample variation across epochs:} \\
    Because augmentation is performed on-the-fly and randomly for each sample, the augmented versions differ each epoch. This continual variation prevents the model from overfitting to a fixed augmented dataset and enhances its exposure to a wider range of feature perturbations.
    \item \textbf{Joint optimization of related augmented samples:} \\
    All augmented views originating from the same input are simultaneously passed through the network during training. This enables AMAuT to recognize that these views share a common semantic source but contain different biases, allowing the model to learn invariances more effectively. As a result, the model gains improved robustness against real-world distortions such as time shifts, Gaussian noise, and background interference.
\end{enumerate}
Together, these two advantages enhance AMAuT’s generalization ability over offline augmentation methods.

For online augmentation methods, as shown in Table~\ref{tab:seq_mlt_cmp}, the proposed multiview approach achieved \textbf{higher test accuracies} on both datasets, for example, an increase from 65.37\% to 68.87\% on CS and from 95.26\% to 95.93\% on SC1, demonstrating that real-time, multiview training substantially improves performance without additional pre-training or data storage requirements.

\noindent\textit{Consistency Analysis}:

To further verify that augmentation-driven multiview learning improves \textbf{convergence stability} and \textbf{generalization}, a \textbf{consistency analysis} was conducted using the \textit{agreement rate metric} (see Eq.~\ref{eq:agreement_rate} in Appendix~\ref{app:cons_analy}).

\begin{table}[ht]
    \centering
    \caption{Comparison of agreement rates between one and four augmentation-driven multiview configurations for AMAuT across five benchmark datasets before applying $TTA^2$.}
    \label{tab:cons_analy}
    \begin{threeparttable}
    \begin{tabular*}{.5\textwidth}{@{\extracolsep\fill}lccc}
        \toprule
        Set & Length & 1-Aug & 4-Aug \\
        \midrule
        CS & 10 s & 0.7100 & 0.7349$\uparrow$ \\
        VS & 12 s & 0.9142 & 0.9231$\uparrow$ \\
        SC1 & 1 s & 0.9592 & 0.9663$\uparrow$ \\
        SC2 & 1 s & 0.9582 & 0.9673$\uparrow$ \\
        AM & 1 s & 0.9995 & N/A \\
        \bottomrule
    \end{tabular*}
    \begin{tablenotes}
        \item[1] “1-Aug” and “4-Aug” indicate the use of one or four augmented views, respectively.
        \item[2] Agreement rate ranges from 0 to 1, where 1 denotes perfect consistency.
        \item[3] For training AM, multiple views are unnecessary, as a single view already achieves above 99.25\% accuracy (see Table~\ref{tab:TTA_refine}). 
    \end{tablenotes}
    \end{threeparttable}
\end{table}
As shown in Table~\ref{tab:cons_analy}, the four-view configuration consistently yielded higher agreement rates across CS, VS, SC1, and SC2. This improvement indicates that augmentation-driven multiview learning enhances model stability and convergence reliability. The results thus provide empirical evidence that multiview augmentation effectively strengthens the robustness and reproducibility of the AMAuT framework. 

\subsubsection*{CNN Kernel Design}
In many existing CNN-based audio classification architectures, relatively small convolutional kernels are commonly used. For example, HuBERT~\cite{hsu2021hubert} and Wav2Vec 2.0~\cite{baevski2020wav2vec} employ bottleneck structures such as \textbf{conv2 + conv3}, while other hybrid CNN-Transformer architectures like ENACT-Heart~\cite{han2025enact} and CNN-TE~\cite{chen2024hybrid} use kernel sizes of \textbf{3, 2, and 1}. These designs typically include only one or two large convolutional layers, such as \textbf{conv14} or \textbf{conv16}, at the head or tail of the CNN stack, rather than throughout its body.

In contrast, this study demonstrates that adopting a \textbf{larger kernel size}, specifically \textbf{conv7}, provides a better trade-off between efficiency and performance for both \textbf{1-second} and \textbf{10-second} audio datasets. The conv7 kernel allows AMAuT to capture longer temporal dependencies without significantly increasing computational complexity. 

\begin{table}[h]
    \centering
    \caption{Performance comparison between conv3 and conv7 kernel structures on CS and SC1 before applying $(TTA)^2$.}
    \label{tab:conv_cmp}
    \begin{threeparttable}
    \begin{tabular*}{.5\textwidth}{@{\extracolsep\fill}ccccl}
        \toprule
        Set & Length & Structure & 1-Aug(\%) & 4-Aug(\%) \\
        \midrule
        CS & 10 s & conv7 & \textbf{65.37} & \textbf{68.87}$\uparrow$ \\
        CS & 10 s & conv3 & 47.57 & 55.98$\uparrow$ \\
        \midrule
        SC1 & 1 s & conv7 & 95.26 & 95.93$\uparrow$ \\
        SC1 & 1 s & conv3 & \textbf{95.39} & \textbf{96.18}$\uparrow$ \\
        \bottomrule
    \end{tabular*}
    \begin{tablenotes}
        \item[1] “Length” indicates audio length in seconds.
        \item[2] “1-Aug” / “4-Aug” denotes the number of augmented views used during testing.
        \item[3] Evaluated on a single RTX 4090 GPU
    \end{tablenotes}
    \end{threeparttable}
\end{table}
As presented in Table~\ref{tab:conv_cmp}, the results confirm that the \textbf{conv7} structure consistently outperforms the \textbf{conv3} configuration before applying $TTA^2$. On the CS with \textbf{10-second} audio, the accuracy improved from 47.57\% (conv3) to \textbf{65.37\%} (conv7) for single-view augmentation, and from 55.98\% to \textbf{68.87\%} when using four augmented views. For SC1 with \textbf{1-second} inputs, the conv7 kernel achieved \textit{95.93\%} compared to \textbf{96.18\%} from the smaller conv3 kernel. Although the performance difference for SC1 was marginal, the results show that conv7 maintains or improves accuracy while providing better temporal representation for longer audio sequences.

Overall, these results suggest that \textbf{medium-sized kernels} (such as conv7) provide a good balance between short and long audio lengths, yielding stable performance for the 1D CNN bottleneck in AMAuT.

\subsubsection*{Contributions of $TTA^2$}
The four refinement strategies, such as Augmentation (Aug), Multi-Training (Mlt), Hybrid (Hyb), and Test-Time Domain Adaptation (TTDA), each aim to improve AMAuT’s inference robustness and overall testing accuracy. These techniques operate without additional labeled data and are designed to enhance the model’s generalization to unseen domains.

\begin{table}[h]
    \centering
    \caption{Test accuracy improvement (\%) from different refinement strategies, including Aug, Mlt, Hyb, and TTDA, applied to AMAuT.}
    \label{tab:TTA_refine}
    \begin{threeparttable}
    \begin{tabular*}{.5\textwidth}{@{\extracolsep\fill}cccccl}
        \toprule
        Set & Orig & Aug & Mlt & Hyb & TTDA \\
        \midrule
        SC1 & 95.93 & 96.02$\uparrow$ & 96.21$\uparrow$ & \textbf{96.31}$\uparrow$ & 96.27$\uparrow$ \\
        SC2 & 95.76 & 95.80$\uparrow$ & \textbf{96.21}$\uparrow$ & 96.13$\uparrow$ & 96.13$\uparrow$ \\
        AM0 & 99.93 & \textbf{99.97}$\uparrow$ & 99.93 & 99.95$\uparrow$ & 99.93 \\
        AM1 & 99.75 & 99.71$\downarrow$ & 99.80$\uparrow$ & 99.80$\uparrow$ & \textbf{99.92}$\uparrow$ \\
        AM2 & \textbf{99.97} & 99.90$\downarrow$ & 99.93$\downarrow$ & 99.91$\downarrow$ & \textbf{99.97} \\
        AM3 & 99.68 & 99.63$\downarrow$ & 99.70$\uparrow$ & 99.68 & \textbf{99.77}$\uparrow$ \\
        AM4 & 99.25 & 99.07$\downarrow$ & 99.38$\uparrow$ & 98.73$\downarrow$ & \textbf{99.52}$\uparrow$ \\
        VS & 91.06 & 90.28$\downarrow$ & \textbf{92.26}$\uparrow$ & 91.56$\uparrow$ & 91.26$\uparrow$ \\
        CS & 68.87 & 69.62$\uparrow$ & 72.46$\uparrow$ & \textbf{72.58}$\uparrow$ & 72.53$\uparrow$ \\
        \bottomrule
    \end{tabular*}
    \begin{tablenotes}
        \item[1] For AudioMNIST (AM), results are reported over five folds (AM0–AM4).
    \end{tablenotes}
    \end{threeparttable}
\end{table}
Experimental results (Table~\ref{tab:TTA_refine}) show that all four methods improved test performance on datasets such as SC1, SC2, VS, and CS. For instance, in SC1, accuracy increased from 95.93\% to \textbf{96.31\%} with the Hybrid method, while TTDA achieved 96.27\%, showing that both ensemble-style and adaptive refinement approaches are effective. Similar improvements were observed for SC2, with accuracy rising from 95.76\% to \textbf{96.21\%} using Mlt and 96.13\% with TTDA.

For the AudioMNIST (AM) dataset, the effects of the four refinement methods were not entirely consistent. While TTDA and Mlt achieved slightly higher cross-validation accuracies (\textbf{99.82\%} and 99.75\%, respectively) compared to the baseline (99.71\%), the Aug and Hyb methods produced minor decreases (99.66\% and 99.45\%, respectively).

Despite these variations, the results demonstrate that applying $TTA^2$ can still yield incremental benefits depending on dataset characteristics and data distributions.

\begin{table}[h]
    \centering
    \caption{Adaptation time comparison for different $TTA^2$ methods in AMAuT using a single RTX 4090 GPU.}
    \label{tab:TTA_cost}
    \begin{threeparttable}
    \begin{tabular*}{.5\textwidth}{@{\extracolsep\fill}ccccl}
        \toprule
        Set & Aug & Mlt & Hyb & TTDA-Epoch \\
        \midrule
        SC1 & \textbf{8 s} & \textbf{8 s} & 24 s & 16 m-50 \\
        SC2 & \textbf{11 s} & \textbf{11 s} & 31 s & 40 m-80 \\
        AM & 8s & \textbf{7 s} & 20 s & 14 m-50 \\
        VS & 9s & \textbf{8 s} & 20 s & 16 m-50 \\
        CS & 64s & \textbf{23 s} & 65 s & 59 m-50 \\
        \bottomrule
    \end{tabular*}
    \begin{tablenotes}
        \item[1] Time is shown in seconds (s) or minutes (m) per method.
        \item[2] “TTDA-Epoch” denotes time costing and the number of epochs used for performing TTDA.
    \end{tablenotes}
    \end{threeparttable}
\end{table}
In terms of computational efficiency, the Mlt method required the least inference time, between \textbf{7} and \textbf{23 seconds}, as reported in Table~\ref{tab:TTA_cost}. By comparison, TTDA was more time-consuming because it involves iterative adaptation across multiple epochs (e.g., \textbf{50–80 epochs}, lasting \textbf{14–59 minutes}, depending on the dataset).

In summary, the results confirm that all four $TTA^2$ strategies can positively impact AMAuT’s generalization at inference. Their effectiveness varies across datasets, suggesting that dataset characteristics, computational constraints, and desired inference efficiency should guide the choice of refinement method. These findings also highlight a promising direction for future research: developing adaptive strategies to automatically select $TTA^2$ methods for optimal performance across different domain settings.

\subsubsection*{CLS+TAL Tokens}
In AMAuT, the \textbf{CLS (class)} and \textbf{TAL (tail)} tokens are positioned at opposite ends of the token sequence (Similar to DeiT~\cite{touvron2021training}).
\begin{table}[h]
    \centering
    \caption{Comparing different positional embedding strategies on CS and VS before applying $TTA^2$.}
    \label{tab:cls_token_cmp}
    \begin{threeparttable}
    \begin{tabular*}{.5\textwidth}{@{\extracolsep\fill}cccl}
        \toprule
        Set & Length & Method & Accuracy (\%) \\
        \midrule
        CS & 10 s & CLS+TAL & \textbf{68.87} \\
        CS & 10 s & CLS & 67.13 \\
        \midrule
        VS & 12 s & CLS+TAL & \textbf{91.06} \\
        VS & 12 s & CLS & 90.56 \\
        \bottomrule
    \end{tabular*}
    \begin{tablenotes}
        \item[1] See CLS token in~\cite{dosovitskiy2020image,touvron2021training}.
        \item[2] 'Length' denotes audio length in seconds.
    \end{tablenotes}
    \end{threeparttable}
\end{table}
The results summarized in Table~\ref{tab:cls_token_cmp} show that CLS+TAl tokens perform better than a single CLS token.

\subsection{Computational Efficiency}
By removing the dependence on pre-training, AMAuT achieves remarkable computational efficiency. The training of AMAuT can be completed entirely on a \textbf{single NVIDIA RTX 4090 GPU}, while maintaining performance comparable to or better than several leading pre-trained models such as SSAST~\cite{gong2022ssast}, EAT~\cite{chen2024eat}, HuBERT~\cite{hsu2021hubert,la2024benchmarking}, and Qwen-Audio~\cite{chu2023qwen} (see Table~\ref{tab:gpu_hour}).

\begin{table}[h]
    \centering
    \caption{Comparison of pre-training and fine-tuning hours for AMAuT and state-of-the-art pre-trained models across multiple datasets.}
    \label{tab:gpu_hour}
    \begin{threeparttable}
    \begin{tabular*}{.5\textwidth}{@{\extracolsep\fill}cccl}
        \toprule
        Algorithm & Set & Pre-train & Fine-Tuning \\
        \midrule
        AMAuT & SC1 & N/A & 4 h 10 m \\
        SSAST~\cite{gong2022ssast} & SC1 & 10 days~\cite{gong2022ssast} & unreported \\
        \midrule
        AMAuT & SC2 & N/A & 5 h 38 m\\
        EAT~\cite{chen2024eat} & SC2 & 58h~\cite{chen2024eat} & unreported \\
        AST~\cite{gong2021ast} & SC2 & 3 days~\cite{touvron2021training} & unreported \\
        PDC~\cite{chrysos2022augmenting} & SC2 & unreported & unreported \\
        \midrule
        AMAuT & AM & N/A & 26 m \\
        HuBERT~\cite{hsu2021hubert,la2024benchmarking} & AM & 137.75 h~\cite{hsu2021hubert} & unreported \\
        XLS-R~\cite{babu2021xls,la2024benchmarking} & AM & unreported & unreported \\
        Wav2Vec2~\cite{baevski2020wav2vec, la2024benchmarking} & AM & 7.5 days~\cite{baevski2020wav2vec} & unreported \\
        \midrule
        AMAuT & VS & N/A & 1 h 28 m \\
        Qwen-Audio~\cite{chu2023qwen} & VS & unreported & unreported \\
        EfficientNet~\cite{gong_vocalsound,tan2019efficientnet} & VS & unknown & unreported \\
        \midrule
        AMAuT & CS & N/A & 4 h 9 m \\
        Audio Flamingo~\cite{kong2024audio} & CS & unreported & unreported \\
        Qwen-Audio~\cite{chu2023qwen} & CS & unreported & unreported \\
        \bottomrule
    \end{tabular*}
    \begin{tablenotes}
        \item[1] Pre-training and fine-tuning durations are reported for each algorithm, where available, from their respective original publications.
    \end{tablenotes}
    \end{threeparttable}
\end{table} 

As shown in Table~\ref{tab:gpu_hour}, SSAST~\cite{gong2022ssast} requires \textbf{four NVIDIA GTX Titan X GPUs}, 1.95 million AudioSet~\cite{gemmeke2017audio} samples, and 281 thousand LibriSpeech~\cite{panayotov2015librispeech} samples for pre-training, a process that takes approximately \textbf{10 days}. EAT~\cite{chen2024eat} performs pre-training on \textbf{four RTX 3090 GPUs} using the AudioSet~\cite{gemmeke2017audio}, requiring about \textbf{58 hours}. AST~\cite{gong2021ast} does not perform self-pre-training but instead reuses weights from DeiT~\cite{touvron2021training}, which itself was trained on ImageNet~\cite{ILSVRC15} using \textbf{four 16 GB V100 GPUs} for \textbf{three days}. HuBERT~\cite{hsu2021hubert,la2024benchmarking} needs \textbf{32 GPUs} and a total of 60,960 hours of audio (960 hours from LibriSpeech~\cite{panayotov2015librispeech} plus 60,000 hours from Libri-Light~\cite{kahn2020libri}) for pre-training, which takes approximately \textbf{137.75 hours}. Similarly, XLS-R~\cite{babu2021xls,la2024benchmarking} requires \textbf{128 GPUs} and several large-scale multilingual datasets, including VoxPopuli~\cite{wang2021voxpopuli}, Multilingual LibriSpeech~\cite{pratap2020mls}, CommonVoice~\cite{ardila2019common}, VoxLingua107~\cite{valk2021voxlingua107}, and BABEL~\cite{gales2014speech}.
Wav2Vec 2.0~\cite{baevski2020wav2vec, la2024benchmarking} also needs \textbf{128 V100 GPUs}, running for \textbf{2.3 days} on LibriSpeech~\cite{panayotov2015librispeech} and \textbf{5.2 days} on Libri-Light~\cite{kahn2020libri}.

Although the pre-training durations and hardware requirements of Qwen-Audio~\cite{chu2023qwen} are not explicitly reported, its large-scale use of \textbf{30 audio datasets} (covering speech, sound, and music) suggests that its pre-training would also be highly resource-intensive. Likewise, Audio Flamingo~\cite{kong2024audio} indicates the use of \textbf{eight NVIDIA A100 GPUs} for pre-training but does not disclose the total compute time or energy cost.

\begin{table}[h]
    \centering
    \caption{Comparison of GPU hours for AMAuT and state-of-the-art pre-trained models across multiple datasets.}
    \label{tab:compute_cost}
    \begin{threeparttable}
    \begin{tabular*}{.5\textwidth}{@{\extracolsep\fill}ccccl}
        \toprule
        Set & AMAuT & Baseline & Baseline-hour & Ratio (\%) \\
        \midrule
        SC1 & 4 h 10 m & SSAST & $4 \times 10$ days & 0.43 \\
        SC2 & 5 h 38 m & EAT & $4\times58$ h & 2.43 \\
        SC2 & 5 h 38 m & AST & $4\times3$ days & 1.96 \\
        AM & 26 m & HuBERT & $32 \times 137.75$ h & 0.01 \\
        AM & 26 m & Wav2Vec2 & $128 \times 7.5$ days & $<0.01$ \\
        \bottomrule
    \end{tabular*}
    \begin{tablenotes}
        \item[1] GPU hours = number of GPUs $\times$ (pre-training + training time). 
        \item[2] Pre-training and fine-tuning durations are reported for each algorithm where available.
        \item[3] Data are collected from original publications unless otherwise noted.
    \end{tablenotes}
    \end{threeparttable}
\end{table}

For GPU hour comparison, AMAuT GPU-hours = per-dataset training time on a single RTX 4090. Baseline GPU-hours are computed as (number of GPUs × reported pretraining hours), where both elements are reported in the baseline paper; if one element is missing, ignore it. Therefore, Table~\ref{tab:compute_cost} compares AMAuT fine-tuning/training cost (single GPU) to baseline pretraining GPU-hours as reported in the literature; it does not compare training/pre-training times under identical hardware or consider baseline fine-tuning costs. Across all datasets, AMAuT required no more than \textbf{3\%} of the GPU hours of these pre-trained counterparts (see Table~\ref{tab:compute_cost}).

In summary, AMAuT achieves accuracy comparable to or exceeding that of state-of-the-art pre-trained models, while operating at \textbf{a fraction} of their GPU hours. This efficiency highlights the practicality of training AMAuT from scratch on a single high-end GPU, making it particularly attractive for researchers and practitioners with limited computational resources.

\section{Discussion}
This section discusses the major insights drawn from the study.

\subsection*{Efficiency}
The most striking outcome is that AMAuT achieves comparable accuracy to the state-of-the-art across all five benchmarks while requiring less than 3\% of the GPU hours of pre-trained models. This result indicates that the architectural adaptability of AMAuT can partially substitute for large-scale pre-training, especially in domain-specific applications.

\subsection*{Flexibility}
AMAuT works across sample rates 16–48 kHz and durations 1–12 s without retraining. For 1D CNN, 
\textbf{medium-sized kernels} (such as conv7) provide a good balance between short and long audio lengths, yielding stable performance for the 1D CNN bottleneck in AMAuT.

\subsection*{Robustness}
Augmentation-driven multiview learning substantially improves robustness over traditional sequential augmentation. Because each epoch generates distinct perturbations, the model continually encounters new audio variants, enhancing its generalization to unseen environments. This finding suggests that online multiview augmentation can be a practical alternative to massive data augmentation pipelines.

Further, $TTA^2$ improves mean accuracy by 0.4–2\% depending on datasets.

\section{Limitations and Future Work}
Despite its promising results, several limitations remain:
\begin{enumerate}
    \item \textbf{Dataset Size Sensitivity} \\
    For stable convergence, datasets with fewer than 15,000 samples tend to underperform, suggesting a lower bound for training purely from scratch.
    \item \textbf{Hyperparameter Dependence} \\
    Performance is sensitive to the Mel-spectrogram~\cite{ustubioglu2023mel,hwang2020mel} configuration and TTDA loss parameters (Eq.~\ref{eq:ttl_tta_loss}). The absence of universal tuning guidelines indicates a need for automated hyperparameter search in future work.
    \item \textbf{Inference Latency in TTDA} \\
    While efficient, TTDA still adds 15–60 minutes per dataset during adaptation, limiting real-time deployment (see Table~\ref{tab:TTA_cost}).
\end{enumerate}



\subsubsection*{Future Directions:}
\begin{itemize}
    \item Investigate \textbf{long-sequence modeling} ($>$30 seconds) using hierarchical temporal segmentation to maintain efficiency.
    \item Developing \textbf{hyperparameter search methods} for Mel-spectrogram and TTDA loss.
    \item Enabling effective training on \textbf{small-scale datasets} with fewer than 15,000 samples.
\end{itemize}

\section{Conclusion}
This paper presents AMAuT, an energy-efficient transformer framework for audio classification, designed as a practical alternative to large foundational models. By eliminating the dependency on pre-training, AMAuT overcomes the limitations of fixed sample rates and restricted audio lengths inherent in existing models. To achieve this, this study introduced a novel architecture and training methodology incorporating augmentation-driven multiview learning, horizontal tokenization, a 1D CNN bottleneck, CLS+TAL tokens, and test-time augmentation/adaptation mechanisms ($TTA^2$).

Consequently, AMAuT can be trained from scratch on diverse datasets. Evaluations on five benchmark datasets, such as AM, SC1, SC2, VS, and CS, demonstrate that AMAuT achieves performance comparable to state-of-the-art pre-trained models like SSAST~\cite{gong2022ssast}, EAT~\cite{chen2024eat}, HuBERT~\cite{hsu2021hubert}, Qwen-Audio~\cite{chu2023qwen}, and Audio Flamingo~\cite{kong2024audio}, with test accuracies of 99.82\%, 96.31\%, 96.21\%, 92.26\%, and 72.58\%, respectively. Notably, this performance is achieved using under 3\% of the GPU time required by the compared models. 

AMAuT demonstrates that \textbf{pretrain-free Transformers} can achieve state-of-the-art performance with drastically reduced computational cost. This efficiency opens new possibilities for practical deployment in embedded, real-time, and low-resource environments.


\section*{Acknowledgments}

The authors gratefully acknowledge the use of language-editing tools to enhance readability. All interpretations and conclusions remain solely those of the authors.

\section*{Data and Code Availability}
The AMAuT implementation and experiment scripts are publicly available at \url{https://github.com/Andy-Shao/AMAuT} to support full reproducibility of results.

\begin{small}
    \bibliographystyle{plain}
    \bibliography{reference}

@String{Computing = "Computing" }

@String{Computer = "{IEEE} Computer" }

@String{Springer = "Springer-Verlag" }

@article{hsu2021hubert,
  title={Hubert: Self-supervised speech representation learning by masked prediction of hidden units},
  author={Hsu, Wei-Ning and Bolte, Benjamin and Tsai, Yao-Hung Hubert and Lakhotia, Kushal and Salakhutdinov, Ruslan and Mohamed, Abdelrahman},
  journal={IEEE/ACM transactions on audio, speech, and language processing},
  volume={29},
  pages={3451--3460},
  year={2021},
  publisher={IEEE}
}

@inproceedings{la2024benchmarking,
  title={Benchmarking representations for speech, music, and acoustic events},
  author={La Quatra, Moreno and Koudounas, Alkis and Vaiani, Lorenzo and Baralis, Elena and Cagliero, Luca and Garza, Paolo and Siniscalchi, Sabato Marco},
  booktitle={2024 IEEE International Conference on Acoustics, Speech, and Signal Processing Workshops (ICASSPW)},
  pages={505--509},
  year={2024},
  organization={IEEE}
}

@inproceedings{gong2022ssast,
  title={Ssast: Self-supervised audio spectrogram transformer},
  author={Gong, Yuan and Lai, Cheng-I and Chung, Yu-An and Glass, James},
  booktitle={Proceedings of the AAAI Conference on Artificial Intelligence},
  volume={36},
  number={10},
  pages={10699--10709},
  year={2022}
}

@inproceedings{chen2024eat,
  title     = {EAT: Self-Supervised Pre-Training with Efficient Audio Transformer},
  author    = {Chen, Wenxi and Liang, Yuzhe and Ma, Ziyang and Zheng, Zhisheng and Chen, Xie},
  booktitle = {Proceedings of the Thirty-Third International Joint Conference on
               Artificial Intelligence, {IJCAI-24}},
  publisher = {International Joint Conferences on Artificial Intelligence Organization},
  editor    = {Kate Larson},
  pages     = {3807--3815},
  year      = {2024},
  month     = {8},
  note      = {Main Track},
  doi       = {10.24963/ijcai.2024/421},
  url       = {https://doi.org/10.24963/ijcai.2024/421},
}

@article{chu2023qwen,
  title={Qwen-audio: Advancing universal audio understanding via unified large-scale audio-language models},
  author={Chu, Yunfei and Xu, Jin and Zhou, Xiaohuan and Yang, Qian and Zhang, Shiliang and Yan, Zhijie and Zhou, Chang and Zhou, Jingren},
  journal={arXiv preprint arXiv:2311.07919},
  year={2023}
}

@article{audiomnist2023,
    title = {AudioMNIST: Exploring Explainable Artificial Intelligence for audio analysis on a simple benchmark},
    journal = {Journal of the Franklin Institute},
    year = {2023},
    issn = {0016-0032},
    doi = {https://doi.org/10.1016/j.jfranklin.2023.11.038},
    url = {https://www.sciencedirect.com/science/article/pii/S0016003223007536},
    author = {Sören Becker and Johanna Vielhaben and Marcel Ackermann and Klaus-Robert Müller and Sebastian Lapuschkin and Wojciech Samek},
}

@INPROCEEDINGS{gong_vocalsound,
  author={Gong, Yuan and Yu, Jin and Glass, James},
  booktitle={ICASSP 2022 - 2022 IEEE International Conference on Acoustics, Speech and Signal Processing (ICASSP)}, 
  title={Vocalsound: A Dataset for Improving Human Vocal Sounds Recognition}, 
  year={2022},
  pages={151-155},
  doi={10.1109/ICASSP43922.2022.9746828}}

@inproceedings{wang2021tent,
  title={Tent: Fully Test-Time Adaptation by Entropy Minimization},
  author={Wang, Dequan and Shelhamer, Evan and Liu, Shaoteng and Olshausen, Bruno and Darrell, Trevor},
  booktitle={International Conference on Learning Representations},
  year={2021},
  url={https://openreview.net/forum?id=uXl3bZLkr3c}
}

@inproceedings{kumar2023conmix,
  title={Conmix for source-free single and multi-target domain adaptation},
  author={Kumar, Vikash and Lal, Rohit and Patil, Himanshu and Chakraborty, Anirban},
  booktitle={Proceedings of the IEEE/CVF Winter Conference on Applications of Computer Vision},
  pages={4178--4188},
  year={2023}
}

@article{liang2024comprehensive,
  title={A comprehensive survey on test-time adaptation under distribution shifts},
  author={Liang, Jian and He, Ran and Tan, Tieniu},
  journal={International Journal of Computer Vision},
  pages={1--34},
  year={2024},
  publisher={Springer}
}

@article{dosovitskiy2020image,
  title={An image is worth 16x16 words: Transformers for image recognition at scale},
  author={Dosovitskiy, Alexey},
  journal={arXiv preprint arXiv:2010.11929},
  year={2020}
}

@article{vaswani2017attention,
  title={Attention is all you need},
  author={Vaswani, A},
  journal={Advances in Neural Information Processing Systems},
  year={2017}
}

@article{ustubioglu2023mel,
  title={Mel spectrogram-based audio forgery detection using CNN},
  author={Ustubioglu, Arda and Ustubioglu, Beste and Ulutas, Guzin},
  journal={Signal, Image and Video Processing},
  volume={17},
  number={5},
  pages={2211--2219},
  year={2023},
  publisher={Springer}
}

@article{hwang2020mel,
  title={Mel-spectrogram augmentation for sequence to sequence voice conversion},
  author={Hwang, Yeongtae and Cho, Hyemin and Yang, Hongsun and Won, Dong-Ok and Oh, Insoo and Lee, Seong-Whan},
  journal={arXiv preprint arXiv:2001.01401},
  year={2020}
}

@article{gong2021ast,
  title={Ast: Audio spectrogram transformer},
  author={Gong, Yuan and Chung, Yu-An and Glass, James},
  journal={arXiv preprint arXiv:2104.01778},
  year={2021}
}

@inproceedings{zhang2022spectrogram,
  title={Spectrogram transformers for audio classification},
  author={Zhang, Yixiao and Li, Baihua and Fang, Hui and Meng, Qinggang},
  booktitle={2022 IEEE International Conference on Imaging Systems and Techniques (IST)},
  pages={1--6},
  year={2022},
  organization={IEEE}
}

@article{baevski2020wav2vec,
  title={wav2vec 2.0: A framework for self-supervised learning of speech representations},
  author={Baevski, Alexei and Zhou, Yuhao and Mohamed, Abdelrahman and Auli, Michael},
  journal={Advances in neural information processing systems},
  volume={33},
  pages={12449--12460},
  year={2020}
}

@inproceedings{panayotov2015librispeech,
  title={Librispeech: an asr corpus based on public domain audio books},
  author={Panayotov, Vassil and Chen, Guoguo and Povey, Daniel and Khudanpur, Sanjeev},
  booktitle={2015 IEEE international conference on acoustics, speech and signal processing (ICASSP)},
  pages={5206--5210},
  year={2015},
  organization={IEEE}
}

@inproceedings{kahn2020libri,
  title={Libri-light: A benchmark for asr with limited or no supervision},
  author={Kahn, Jacob and Riviere, Morgane and Zheng, Weiyi and Kharitonov, Evgeny and Xu, Qiantong and Mazar{\'e}, Pierre-Emmanuel and Karadayi, Julien and Liptchinsky, Vitaliy and Collobert, Ronan and Fuegen, Christian and others},
  booktitle={ICASSP 2020-2020 IEEE International Conference on Acoustics, Speech and Signal Processing (ICASSP)},
  pages={7669--7673},
  year={2020},
  organization={IEEE}
}

@inproceedings{gemmeke2017audio,
  title={Audio set: An ontology and human-labeled dataset for audio events},
  author={Gemmeke, Jort F and Ellis, Daniel PW and Freedman, Dylan and Jansen, Aren and Lawrence, Wade and Moore, R Channing and Plakal, Manoj and Ritter, Marvin},
  booktitle={2017 IEEE international conference on acoustics, speech and signal processing (ICASSP)},
  pages={776--780},
  year={2017},
  organization={IEEE}
}

@inproceedings{touvron2021training,
  title={Training data-efficient image transformers \& distillation through attention},
  author={Touvron, Hugo and Cord, Matthieu and Douze, Matthijs and Massa, Francisco and Sablayrolles, Alexandre and J{\'e}gou, Herv{\'e}},
  booktitle={International conference on machine learning},
  pages={10347--10357},
  year={2021},
  organization={PMLR}
}

@inproceedings{deng2009imagenet,
  title={Imagenet: A large-scale hierarchical image database},
  author={Deng, Jia and Dong, Wei and Socher, Richard and Li, Li-Jia and Li, Kai and Fei-Fei, Li},
  booktitle={2009 IEEE conference on computer vision and pattern recognition},
  pages={248--255},
  year={2009},
  organization={Ieee}
}

@inproceedings{chrysos2022augmenting,
  title={Augmenting deep classifiers with polynomial neural networks},
  author={Chrysos, Grigorios G and Georgopoulos, Markos and Deng, Jiankang and Kossaifi, Jean and Panagakis, Yannis and Anandkumar, Anima},
  booktitle={European Conference on Computer Vision},
  pages={692--716},
  year={2022},
  organization={Springer}
}

@article{babu2021xls,
  title={XLS-R: Self-supervised cross-lingual speech representation learning at scale},
  author={Babu, Arun and Wang, Changhan and Tjandra, Andros and Lakhotia, Kushal and Xu, Qiantong and Goyal, Naman and Singh, Kritika and Von Platen, Patrick and Saraf, Yatharth and Pino, Juan and others},
  journal={arXiv preprint arXiv:2111.09296},
  year={2021}
}

@article{ILSVRC15,
Author = {Olga Russakovsky and Jia Deng and Hao Su and Jonathan Krause and Sanjeev Satheesh and Sean Ma and Zhiheng Huang and Andrej Karpathy and Aditya Khosla and Michael Bernstein and Alexander C. Berg and Li Fei-Fei},
Title = {{ImageNet Large Scale Visual Recognition Challenge}},
Year = {2015},
journal   = {International Journal of Computer Vision (IJCV)},
doi = {10.1007/s11263-015-0816-y},
volume={115},
number={3},
pages={211-252}
}

@article{wang2021voxpopuli,
  title={VoxPopuli: A large-scale multilingual speech corpus for representation learning, semi-supervised learning and interpretation},
  author={Wang, Changhan and Riviere, Morgane and Lee, Ann and Wu, Anne and Talnikar, Chaitanya and Haziza, Daniel and Williamson, Mary and Pino, Juan and Dupoux, Emmanuel},
  journal={arXiv preprint arXiv:2101.00390},
  year={2021}
}

@article{pratap2020mls,
  title={Mls: A large-scale multilingual dataset for speech research},
  author={Pratap, Vineel and Xu, Qiantong and Sriram, Anuroop and Synnaeve, Gabriel and Collobert, Ronan},
  journal={arXiv preprint arXiv:2012.03411},
  year={2020}
}

@article{ardila2019common,
  title={Common voice: A massively-multilingual speech corpus},
  author={Ardila, Rosana and Branson, Megan and Davis, Kelly and Henretty, Michael and Kohler, Michael and Meyer, Josh and Morais, Reuben and Saunders, Lindsay and Tyers, Francis M and Weber, Gregor},
  journal={arXiv preprint arXiv:1912.06670},
  year={2019}
}

@inproceedings{valk2021voxlingua107,
  title={VoxLingua107: a dataset for spoken language recognition},
  author={Valk, J{\"o}rgen and Alum{\"a}e, Tanel},
  booktitle={2021 IEEE Spoken Language Technology Workshop (SLT)},
  pages={652--658},
  year={2021},
  organization={IEEE}
}

@inproceedings{gales2014speech,
  title={Speech recognition and keyword spotting for low-resource languages: Babel project research at cued},
  author={Gales, Mark JF and Knill, Kate M and Ragni, Anton and Rath, Shakti P},
  booktitle={Fourth International workshop on spoken language technologies for under-resourced languages (SLTU-2014)},
  pages={16--23},
  year={2014},
  organization={International Speech Communication Association (ISCA)}
}

@inproceedings{tan2019efficientnet,
  title={Efficientnet: Rethinking model scaling for convolutional neural networks},
  author={Tan, Mingxing and Le, Quoc},
  booktitle={International conference on machine learning},
  pages={6105--6114},
  year={2019},
  organization={PMLR}
}

@inproceedings{szegedy2016rethinking,
  title={Rethinking the inception architecture for computer vision},
  author={Szegedy, Christian and Vanhoucke, Vincent and Ioffe, Sergey and Shlens, Jon and Wojna, Zbigniew},
  booktitle={Proceedings of the IEEE conference on computer vision and pattern recognition},
  pages={2818--2826},
  year={2016}
}

@inproceedings{devlin2019bert,
  title={Bert: Pre-training of deep bidirectional transformers for language understanding},
  author={Devlin, Jacob and Chang, Ming-Wei and Lee, Kenton and Toutanova, Kristina},
  booktitle={Proceedings of the 2019 conference of the North American chapter of the association for computational linguistics: human language technologies, volume 1 (long and short papers)},
  pages={4171--4186},
  year={2019}
}

@article{warden2018speech,
  title={Speech commands: A dataset for limited-vocabulary speech recognition},
  author={Warden, Pete},
  journal={arXiv preprint arXiv:1804.03209},
  year={2018}
}

@article{kong2024audio,
  title={Audio flamingo: A novel audio language model with few-shot learning and dialogue abilities},
  author={Kong, Zhifeng and Goel, Arushi and Badlani, Rohan and Ping, Wei and Valle, Rafael and Catanzaro, Bryan},
  journal={arXiv preprint arXiv:2402.01831},
  year={2024}
}

@inproceedings{jeong2022cochlscene,
  title={Cochlscene: Acquisition of acoustic scene data using crowdsourcing},
  author={Jeong, Il-Young and Park, Jeongsoo},
  booktitle={2022 Asia-Pacific Signal and Information Processing Association Annual Summit and Conference (APSIPA ASC)},
  pages={17--21},
  year={2022},
  organization={IEEE}
}

@article{kim2023sgem,
  title={Sgem: Test-time adaptation for automatic speech recognition via sequential-level generalized entropy minimization},
  author={Kim, Changhun and Park, Joonhyung and Shim, Hajin and Yang, Eunho},
  journal={arXiv preprint arXiv:2306.01981},
  year={2023}
}

@article{loo2025temporal,
  title={Temporal patterns in Malaysian rainforest soundscapes demonstrated using acoustic indices and deep embeddings trained on time-of-day estimation},
  author={Loo, Yen Yi and Lee, Mei Yi and Shaheed, Samien and Maul, Tomas and Clink, Dena Jane},
  journal={The Journal of the Acoustical Society of America},
  volume={157},
  number={1},
  pages={1--16},
  year={2025},
  publisher={AIP Publishing}
}

@book{volkenstein2009entropy,
  title={Entropy and information},
  author={Volkenstein, Mikhail V},
  volume={57},
  year={2009},
  publisher={Springer Science \& Business Media}
}

@inproceedings{kimura2021understanding,
  title={Understanding test-time augmentation},
  author={Kimura, Masanari},
  booktitle={International Conference on Neural Information Processing},
  pages={558--569},
  year={2021},
  organization={Springer}
}

@article{han2025enact,
  title={ENACT-Heart--ENsemble-based Assessment Using CNN and Transformer on Heart Sounds},
  author={Han, Jiho and Shaout, Adnan},
  journal={arXiv preprint arXiv:2502.16914},
  year={2025}
}

@article{chen2024hybrid,
  title={A hybrid parallel computing architecture based on CNN and transformer for music genre classification},
  author={Chen, Jiyang and Ma, Xiaohong and Li, Shikuan and Ma, Sile and Zhang, Zhizheng and Ma, Xiaojing},
  journal={Electronics},
  volume={13},
  number={16},
  pages={3313},
  year={2024},
  publisher={MDPI}
}
\end{small}

\appendix
\section{Algorithm Details}\label{app:algorithm}
\subsection{Multiview Augmentation}
\begin{algorithm}[h]
    \caption{Augmentation-driven Multiview Learning}\label{alg:aug_multiview}
    \begin{algorithmic}[1]
        \REQUIRE ($X, Y$) is the sample features and labels. $lr>0$ is the learning rate.
        \ENSURE $\mathcal{L}$ is the loss function (training loss requires labels, but TTDA loss does not). $\mathcal{U}(\cdot)$ is the uniform distribution. $aug(\cdot)_k$ is the $k$-th augmentation method. 
        \STATE random initialize $\theta$
        \FOR{$X_i, Y_i \sim \mathcal{U}(X,Y)$ }
            \STATE $L \leftarrow 0$
            \FOR{$k \leftarrow 1$ to $K$}
                \STATE $\bar{X}_i \leftarrow aug_k(X_i)$
                \STATE $L \leftarrow L + \mathcal{L}(f(\bar{X}_i; \theta),Y_i)$
            \ENDFOR
            \STATE $\theta \leftarrow \theta - lr\nabla_{\theta}L$
        \ENDFOR
    \end{algorithmic}
\end{algorithm}
\subsection{Test-time Domain Adaptation (TTDA)}
The three components of TTDA are defined as follows:
\begin{itemize}
    \item \textbf{Nuclear-Norm Maximization (NM)}~\cite{kumar2023conmix}: \\
    Encourages diverse class separation.
    \begin{equation}\label{eq:nm_loss}
        \mathcal{L}_{NM} = -\frac{1}{B\times C}||\{f(x_j;\theta) | j \in [1, B]\}||_F
    \end{equation}
    \item \textbf{Entropy Minimization (EN)}~\cite{wang2021tent}: \\
    Generally, a low entropy value indicates a high degree of certainty. Reducing entropy is reducing uncertainty.
    \begin{equation}\label{eq:entropy_loss}
        \mathcal{L}_{EN} = - \frac{1}{B}\sum^B_j \sum_{i}^C \hat{p}_{i,j} \log{(\hat{p}_{i,j} + \epsilon)} 
    \end{equation}
    \item A modified \textbf{Generalized Entropy (GEN)}~\cite{kim2023sgem}: \\
    \begin{equation}\label{eq:g-entropy_loss}
        \mathcal{L}_{GEN} = \frac{1}{B} \sum^B_j \frac{1-\sum_i^C \hat{p}^{q}_{i,j}}{q-1}
    \end{equation}
\end{itemize}
Here, $B$ denotes the batch size, $C$ the number of classes, $q$ is scaling coefficient, $\epsilon=10^{-6}$ avoids division by zero, and $||\cdot||_F$ the Frobenius norm.

\subsection{Test-time Augmentation (TTAu)}
\subsubsection*{(a) Augmentation-driven refinement (Aug)} 
This common TTAu approach generates $A$ (even number) weakly augmented variants of each test input and averages their predictions with that of the original sample:
\begin{equation}\label{eq:aug_rf}
    \begin{split}
        & \text{aug\_rf}(f,x,A;\theta) \\
        & = \frac{1}{A+1}\Big[f(x;\theta) + \sum_{i=1}^A f(\text{aug}^{(i)}(x);\theta)\Big]
    \end{split}
\end{equation}
where $x$ is a test sample, and $aug^{(i)}$ is the $i$-th augmentation function. In experiments, Aug applied three views, such as \emph{(1)} the original sample $x$, \emph{(2-3)} \textbf{left} and \textbf{right time-shift augmentations} ($\pm17\%$), consistent with those used during prediction, to average outputs. All augmentations were executed on the CPU, so processing time depends on the efficiency of CPU-GPU coordination.

\subsubsection*{(b) Multi-training-based refinement (Mlt)}
This strategy averages predictions from multiple independently trained AMAuT models:
\begin{equation}\label{eq:mlt_rf}
    \text{mlt}(x) = \frac{1}{M} \sum_{i=1}^M f(x;\theta^{(i)})
\end{equation}
where $M$ is an odd integer ($M=2n+1, n\in \mathbb{Z}$) and $\theta^{(i)}$ denotes the parameters of the $i$-th trained model. In experiments, $M=3$. This ensemble-like refinement mitigates model-specific biases.

\subsubsection*{(c) Hybrid refinement (Hyb)} 
The hybrid method combines the benefits of both augmentation- and multi-training-based refinements:
\begin{equation}\label{eq:hyb_rf}
    \text{hyb}(x) = \frac{1}{M} \sum_{i=1}^M \text{aug\_rf}(f,x,A;\theta^{(i)})
\end{equation}
In experiments, $M=3$ and $A=2$.

\subsection{Training Loss}
To prevent overconfidence and improve generalization, this study leverages \textbf{cross-entropy} and \textbf{label-smoothing regularization (LSR)}~\cite{szegedy2016rethinking} for training loss:
\begin{equation}\label{eq:train_loss}
    \mathcal{L}_{Tr} = -\frac{1}{B} \sum^B_j \sum^C_i \big[(1-\epsilon)p_{i,j} + \frac{\epsilon}{C}\big] \log{\hat{p}_{i,j}}
\end{equation}
where $B$ is the batch size, $C$ the number of classes, $\epsilon = 0.1$ the smoothing parameter, $p_{i,j}$ the ground-truth probability, and $\hat{p}_{i,j}$ the model’s predicted probability for class $i$ for sample $j$. This formulation improves model stability and reduces overfitting during training.

\section{Consistency Analysis}\label{app:cons_analy}
To evaluate the consistency of independently trained AMAuT models, an agreement rate is computed based on the alignment of their predictions. Let the confusion matrix between two prediction sets $p_1$ and $p_2$ be defined as:
\begin{equation}
    cm = \text{confusion\_matrix}(p_1, p_2)
\end{equation}
The overall agreement rate is then calculated as:
\begin{equation}
    \text{agreement rate} = \frac{\sum_i cm_{i,i}}{\sum_{i,j} cm_{i,j}} \label{eq:agreement_rate}
\end{equation}
where $cm_{i,i}$ denotes the number of samples for which both predict the same class label $i$, and $\sum_{i,j} cm_{i,j}$ represents the total number of samples considered.

A higher agreement rate indicates greater consistency between models trained independently under the same configuration, suggesting stable convergence and robust generalization of the AMAuT framework.

\section{Hyperparameter Settings}\label{app:param_set}
\begin{table}[ht]
    \centering
    \caption{Hyperparameter Settings}
    \label{tab:ttda_param}\label{tab:LRS_param}
    \begin{threeparttable}
    \begin{tabular*}{.5\textwidth}{@{\extracolsep\fill}lccccccc}
        \toprule
        Set & $\alpha$ & $\beta$ & $\gamma$ & $q$ & $lr$ & $\lambda$ & $\eta$ \\
        \midrule
        AM & 1.0 & 0.5 & 0.5 & 1.1 & $10^{-3}$ & 10 & 40 \\
        SC1 & 0.2 & 1.0 & 0.0 & 0.8 & $10^{-3}$ & 10 & 50 \\
        SC2 & 0.0 & 1.0 & 0.2 & 1.1 & $10^{-3}$ & 10 & 50 \\
        VS & 1.0 & 0.5 & 0.5 & 1.1 & $10^{-3}$ & 10 & 80 \\
        CS & 1.0 & 0.5 & 0.5 & 1.1 & $10^{-3}$ & 10 & 40 \\
        \bottomrule
    \end{tabular*}
    \begin{tablenotes}
        \item[1] The parameters, $\alpha$, $\beta$, and $\gamma$, correspond to the weighting coefficients in Eq.~\ref{eq:ttl_tta_loss}.
        \item[2] The coefficient $q$ denotes the scaling factor in Eq.~\ref{eq:g-entropy_loss}.
        \item[3] The learning rate $lr$, adjustment factor $\lambda$, and scheduling cardinality $\eta$ follow the update rule defined in Eq.~\ref{eq:lr_scheduler}. 
    \end{tablenotes}
    \end{threeparttable}
\end{table}


\end{document}